\documentclass[aps,twocolumn,showkeys,showpacs]{revtex4}%revtex4-1
\usepackage{amssymb}
\usepackage{amsmath}
\usepackage{amscd}
\usepackage{graphicx}
\usepackage{subfigure}
\usepackage{float}
\usepackage{array}
\begin{document}

\title[Short Title]{Ground state of the asymmetric Rabi model in the ultrastrong coupling regime}

\author{Li-Tuo Shen}
%\email{lituoshen@gmail.com}
\author{Zhen-Biao Yang}
\email{zbyang@fzu.edu.cn}
\author{Mei Lu}
\author{Rong-Xin Chen}
\author{Huai-Zhi Wu}
%\author{Shi-Biao Zheng}
%\email{sbzheng11@163.com}

\affiliation{Lab of Quantum Optics, Department of Physics, Fuzhou
University, Fuzhou 350002, China}
%\begin{spacing}{2.0}

\begin{abstract}
We study the ground states of the single- and two-qubit asymmetric
Rabi models, in which the qubit-oscillator coupling strengths for
the counterrotating-wave and corotating-wave interactions are
unequal. We take the transformation method to obtain the
approximately analytical ground states for both models and
numerically verify its validity for a wide range of parameters under
the near-resonance condition. We find that the ground-state energy
in either the single- or two-qubit asymmetric Rabi model has an
approximately quadratic dependence on the coupling strengths
stemming from different contributions of the counterrotating-wave
and corotating-wave interactions. For both models, we show that the
ground-state energy is mainly contributed by the
counterrotating-wave interaction. Interestingly, for the two-qubit
asymmetric Rabi model, we find that, with the increase of the
coupling strength in the counterrotating-wave or corotating-wave
interaction, the two-qubit entanglement first reaches its maximum
then drops to zero. Furthermore, the maximum of the two-qubit
entanglement in the two-qubit asymmetric Rabi model can be much
larger than that in the two-qubit symmetric Rabi model.
\end{abstract}
 \pacs{42.50.Ct, 42.50.Pq, 03.65.Ud}
  \keywords{ground state, asymmetric coupling, Rabi model, ultrastrong coupling}
\maketitle

\noindent
\section{Introduction}

The Rabi model \cite{PR-49-324-1936}, describing the interaction
between a two-level system and a quantized harmonic oscillator, is a
fundamental model in quantum optics. For the cavity quantum
electrodynamics (QED) experiments, the qubit-oscillator coupling
strength of the Rabi model is far smaller than the oscillator's
frequency and the corotating-wave approximation (RWA) works well,
bringing in the ubiquitous Jaynes-Cummins model
\cite{IEEE-51-89-1963,JMO-40-1195-1993,PRL-87-037902-2001,PRA-71-013817-2005}.
With recent experiment progresses in Rabi models
\cite{PT-58-42-2005,Science-326-108-2009,PR-492-1-2010,RPP-74-104401-2011,Nature-474-589-2011,RMP-84-1-2012,
RMP-85-623-2013,arxiv1308-6253-2014} in the ultrastrong coupling
regime
\cite{PRB-78-180502-2008,PRB-79-201303-2009,Nature-458-178-2009,Nature-6-772-2010,
PRL-105-237001-2010,PRL-105-196402-2010,PRL-106-196405-2011,
Science-335-1323-2012,PRL-108-163601-2012,PRB-86-045408-2012,NatureCommun-4-1420-2013},
in which the qubit-oscillator coupling strength becomes a
considerable fraction of the oscillator's or qubit's frequency, the
RWA breaks down but relatively complex quantum dynamics arises,
bringing about many fascinating quantum phenomena
\cite{NJP-13-073002-2011,PRL-109-193602-2012,PRA-81-042311-2010,PRA-87-013826-2013,
PRA-59-4589-1999,PRA-62-033807-2000,PRB-72-195410-2005,PRA-74-033811-2006,
PRA-77-053808-2008,PRA-82-022119-2010,PRL-107-190402-2011,PRL-108-180401-2012,PLA-376-349-2012}.

Explicitly analytic solution to the Rabi model beyond the RWA is
hard to obtain due to the non-integrability in its
infinite-dimensional Hilbert space. Since it is difficult to capture
the physics through numerical solution
\cite{JPA-29-4035-1996,EPL-96-14003-2011}, various approximately
analytical methods for obtaining the ground states of the symmetric
Rabi models (SRM) have been tried
\cite{RPB-40-11326-1989,PRB-42-6704-1990,PRL-99-173601-2007,EPL-86-54003-2009,
PRA-80-033846-2009,PRL-105-263603-2010,PRA-82-025802-2010,
EPJD-66-1-2012,PRA-86-015803-2012,
PRA-85-043815-2012,PRA-86-023822-2012,EPJB-38-559-2004,
PRB-75-054302-2007,EPJD-59-473-2010,arXiv-1303-3367v2-2013,arXiv-1305-1226-2013,PRA-87-022124-2013,PRA-86-014303-2012,arXiv-1305-6782-2013}.
Especially, Braak \cite{PRL-107-100401-2011} used the method based
on the $Z_{2}$ symmetry to analytically determine the spectrum of
the single-qubit Rabi model, which was dependent on the composite
transcendental function defined through its power series but failed
to derive the concrete form of the system's ground state. In Ref.
\cite{PRA-81-042311-2010}, Ashhab \emph{et al.} applied the method
of adiabatic approximation to treat two extreme situations to obtain
the eigenstates and eigenenergies in the single-qubit SRM, i.e., the
situation with a high-frequency oscillator or a high-frequency
qubit. Ashhab \cite{PRA-87-013826-2013} used different order
parameters to identify the phase regions of the single-qubit SRM and
found that the phase-transition-like behavior appeared when the
oscillator's frequency was much lower than the qubit's frequency.
Lee and Law \cite{arXiv-1303-3367v2-2013} used the transformation
method to seek the approximately analytical ground state of the
two-qubit SRM in the near-resonance regime, and found that the
two-qubit entanglement drops as the coupling strength further
increased after it reached its maximum.

Previous studies consider the ground state of the SRM, i.e., the
qubit-oscillator coupling strengths of the counterrotating-wave and
corotating-wave interactions are equal. In this paper, we study the
asymmetic Rabi models (ASRM), i.e., the coupling strengths for the
counterrotating-wave and corotating-wave interactions are unequal,
which helps to gain deep insight into the fundamentally physical
property of such models. Different from Refs.
\cite{PRA-81-042311-2010,PRA-87-013826-2013}, we here use the
transformation method to obtain the ground state of the single-qubit
ASRM under the near-resonance situation, where the oscillator's
frequency approximates the qubit's frequency. Differ further from
Ref. \cite{arXiv-1303-3367v2-2013}, our investigation for the
two-qubit ASRM intuitively identifies the collective contribution to
its ground-state entanglement caused by the corotating-wave and
counterrotating-wave interactions.

We investigate the single- and two-qubit ASRMs and show that their
approximately analytical ground states agree well with the exactly
numerical solutions for a wide range of parameters under the
near-resonance situation, and the ground-state energy has an
approximately quadratic dependence on the coupling strengths
stemming from contributions of the counterrotating-wave and
corotating-wave interactions. Besides, we show that the ground-state
energy is mainly contributed by the counterrotating-wave interaction
in both models. For the two-qubit ASRM, we obtain the approximately
analytical negativity. Interestingly, for the two-qubit ASRM, we
find that, with the increase of the coupling strength in the
counterrotating-wave or corotating-wave interaction, the two-qubit
entanglement first reaches its maximum then drops to zero.

% Our
%study in this paper mainly reveals the collective contribution of
%the qubit-oscillator coupling strengths of the counterrotating-wave
%and corotating-wave interactions to the ASRM's ground states.

The advantages of our result are the collective contributions to the
ground state of the ASRM caused by the corotating-wave interaction
and counterrotating-wave interaction can be determined
approximately, and the contribution of the counterrotating-wave
interaction on the ground state energy is larger than that of the
corotating-wave interaction. We find that the maximal two-qubit
entanglement of the ASRM is larger than that in the case of SRM.
However, the transformation method here is applicable to the ASRM
only under the near-resonant regime, where the oscillator's
frequency approximates the qubit's frequency. When the
corotating-wave and counterrotating-wave coupling constants are
large enough in the ASRM, the result obtained by the transformation
method has a big error compared with that obtained by the exactly
numerical method. Such an investigation can also be generalized to
the complex cases of three- and more-qubit ASRM. Note that the ASRM
can be realized by using two unbalanced Raman channels between two
atomic ground states induced by a cavity mode and two classical
fields in theory \cite{PRA-75-013804-2007}.

\section{The single-qubit ASRM}

\subsection{Transformed ground state}

The Hamiltonian of the single-qubit ASRM is \cite{PRA-8-1440-1973}:
(assume $\hbar=1$ for simplicity hereafter)
\begin{eqnarray}\label{1}
H_{1}&=&\frac{1}{2}w_{a}\sigma_{z}+w_{b}b^{\dagger}b\cr&&
+\frac{\lambda_{1}}{2}(b^{\dagger}\sigma_{-}+b\sigma_{+})
+\frac{\lambda_{2}}{2}(b^{\dagger}\sigma_{+}+b\sigma_{-}),
\end{eqnarray}
where $w_{a}$ is the qubit's frequency. $\sigma_{z}$ and
$\sigma_{\pm}$ are the Pauli matrices, describing the qubit's energy
operator and the spin-flip operators, respectively. We assume that
$|\downarrow\rangle_{A}$ and $|\uparrow\rangle_{A}$ are the
eigenstates of $\sigma_{z}$, i.e., $\sigma_{z}$
$|\downarrow\rangle_{A}$ $=$ $-|\downarrow\rangle_{A}$ and
$\sigma_{z}$ $|\uparrow\rangle_{A}$ $=$ $|\uparrow\rangle_{A}$.
$b^{\dagger}$ ($b$) is the creation (annihilation) operator of the
harmonic oscillator with the frequency $w_{b}$. The qubit-oscillator
coupling strengths of the corotating-wave interaction
$(b^{\dagger}\sigma_{-}+b\sigma_{+})$ and the counterrotating-wave
interaction $(b^{\dagger}\sigma_{+}+b\sigma_{-})$ are denoted by
$\lambda_{1}$ and $\lambda_{2}$, respectively. However, when
$\lambda_{1}$ $\neq$ $\lambda_{2}$ (here $\lambda_{1}$,
$\lambda_{2}$, $w_{a}$ $\neq0$), to our knowledge, there is still no
analytical solution to the ground state of the single-qubit ASRM.

Our task in this paper is to determine the ground-state energy $E_g$
and the ground-state vector $|\phi_g\rangle$ for the single-
(Section II) or two-qubit (Section III) ASRM, where
$H_{1}|\phi_g\rangle$ $=$ $E_{g}|\phi_g\rangle$. In this paper, the
subscripts $A$ and $F$ denote the vectors of the atomic state and
field state, respectively.

To deal with the counterrotating-wave terms in Eq. (\ref{1}), we
apply a unitary transformation to the Hamiltonian $H_{1}$
\cite{EPJD-59-473-2010,PRB-75-054302-2007,PRA-82-022119-2010}:
\begin{eqnarray}\label{3}
H_{1}^{'}&=&e^{S_{1}}H_{1}e^{-S_{1}},
\end{eqnarray}
with
\begin{eqnarray}\label{4}
S_{1}=\xi_{1}(b^{\dagger}-b)\sigma_{x},
\end{eqnarray}
where $\xi_{1}$ is a variable to be determined later. Then the
transformed Hamiltonian $H_{1}^{'}$ can be decomposed into three
parts:
\begin{eqnarray}\label{5}
H_{1}^{'}&=&H_{1}^{a}+H_{1}^{b}+H_{1}^{c},
\end{eqnarray}
with
\begin{eqnarray}\label{6-7-8}
H_{1}^{a}&=&\frac{1}{2}\big[w_{a}\eta_{1}-(\lambda_{1}-\lambda_{2})\xi_{1}\eta_{1}\big]\sigma_{z}\cr\cr&&+
\big[w_{b}-(\lambda_{1}-\lambda_{2})\xi_{1}\eta_{1}\sigma_{z}\big]b^{\dagger}b\cr\cr&&
+w_{b}\xi_{1}^{2}-\frac{1}{2}(\lambda_{1}+\lambda_{2})\xi_{1},\\
H_{1}^{b}&=&\big[\frac{1}{4}(\lambda_{1}+\lambda_{2})-w_{b}\xi_{1}\big](b^{\dagger}+b)\sigma_{x}\cr\cr&&
-i\big[\frac{1}{4}(\lambda_{1}-\lambda_{2})\eta_{1}+w_{a}\xi_{1}\eta_{1}\big](b^{\dagger}-b)\sigma_{y},\\
H_{1}^{c}&=&\frac{1}{2}w_{a}\sigma_{z}\bigg\{\cosh\big[2\xi_{1}(b^{\dagger}-b)\big]-\eta_{1}
\bigg\}\cr&&
-\frac{i}{2}w_{a}\sigma_{y}\bigg\{\sinh\big[2\xi_{1}(b^{\dagger}-b)\big]-2\xi_{1}\eta_{1}(b^{\dagger}-b)
\bigg\}\cr&&
-\frac{i}{4}(\lambda_{1}-\lambda_{2})(b^{\dagger}-b)\sigma_{y}\bigg\{\cosh\big[2\xi_{1}(b^{\dagger}-b)\big]\cr&&-\eta_{1}
\bigg\}+\frac{1}{4}(\lambda_{1}-\lambda_{2})(b^{\dagger}-b)\sigma_{z}\bigg\{\sinh\big[2\xi_{1}(b^{\dagger}
\cr&&-b)\big]-2\xi_{1}\eta_{1}(b^{\dagger}-b)
\bigg\}+O(b^{\dagger2},b^{2}),
\end{eqnarray}
where $\eta_{1}$$=_{F}$$\langle
0|\cosh[2\xi_{1}(b^{\dagger}-b)]|0\rangle_{F}$ $=$ $e^{-2\xi_{1}^2}$
and
$O(b^{\dagger2},b^{2})=\frac{1}{2}(\lambda_{1}-\lambda_{2})\xi_{1}\eta_{1}(b^{\dagger2}+b^2)\sigma_{z}$.
The terms $\cosh[2\xi_{1}(b^{\dagger}-b)]$ and
$\sinh[2\xi_{1}(b^{\dagger}-b)]$ in $H_{1}^{c}$ have the dominating
expansions \cite{EPJD-59-473-2010}:
\begin{eqnarray}\label{9-10}
\cosh[2\xi_{1}(b^{\dagger}-b)]&\simeq&\eta_{1}+O(\xi_{1}^2),\\
\sinh[2\xi_{1}(b^{\dagger}-b)]&\simeq&2\xi_{1}\eta_{1}(b^{\dagger}-b)+O(\xi_{1}^3),
\end{eqnarray}
where $O(b^{\dagger2},b^{2})$, $O(\xi_{1}^2)$ and $O(\xi_{1}^3)$ are
higher-order terms of $b^{\dagger}$ and $b$, which represent the
double- and three-photon transition processes and can be neglected
as an approximation when $\xi_1$ and $|\lambda_1\pm\lambda_2|$ are
much smaller than the frequency sum $w_a+w_b$ where $w_a\approx
w_b$. Thus, $H_{1}^{'}\simeq H_{1}^{a}+H_{1}^{b}$.

When the parameter $\xi_{1}$ is chosen such that it satisfies the
condition:
\begin{eqnarray}\label{11}
e^{2\xi_{1}^2}\big[(\lambda_{1}+\lambda_{2})-4w_{b}\xi_{1}\big]=(\lambda_{1}-\lambda_{2})+4w_{a}\xi_{1},
\end{eqnarray}
the qubit and the oscillator are coupled in the following form:
\begin{eqnarray}\label{12}
H_{1}^{b}&=&\frac{1}{2}\big[(\lambda_{1}+\lambda_{2})-4w_{b}\xi_{1}\big]\times\big(b^{\dagger}
\sigma_{-}+b\sigma_{+}\big).
\end{eqnarray}
Note that $H_{1}^{b}$ in Eq. (\ref{12}) contains no
counterrotating-wave interactions in which the qubit excitation
(deexcitation) is accompanied by the emission (absorption) of a
photon. Therefore, the transformed Hamiltonian $H_{1}^{'}$ is
exactly solvable when we eliminate the counterrotating-wave terms by
choosing $\xi_{1}$ to satisfy Eq. (\ref{11}) and by neglecting
higher-order transition processes which are presented by terms
$O(b^{\dagger2},b^{2})$, $O(\xi_{1}^2)$ and $O(\xi_{1}^3)$.

It is easy to show that the eigenvector
$|\downarrow\rangle_A|0\rangle_F$ is the ground-state vector of the
transformed Hamiltonian $H_{1}^{'}$, with $|0\rangle_{F}$ being the
vacuum state of the harmonic oscillator, and the corresponding
eigenenergy $E_{g1}$ is:
\begin{eqnarray}\label{13}
E_{g1}=\xi_1^2w_{b}-\frac{1}{2}(\lambda_1+\lambda_2)\xi_1-\frac{1}{2}\eta_1[w_a-\xi_1(\lambda_1-\lambda_2)].\cr&&
\end{eqnarray}
We see that when $\lambda_{1}=\lambda_{2}$, $E_{g1}$ reduces to the
transformed ground-state energy derived in Ref.
\cite{EPJD-59-473-2010}. Therefore, the ground state of the original
Hamiltonian (\ref{1}) can be approximately constructed:
\begin{eqnarray}\label{14}
|\phi_{g1}\rangle&=&e^{-S_{1}}|\downarrow\rangle_{A}|0\rangle_{F}\cr
&=&\frac{1}{\sqrt{2}}(|\psi_{A}^{+}\rangle|-\xi_1\rangle_F-|\psi_{A}^{-}\rangle|\xi_1\rangle_F),
\end{eqnarray}
with $|\xi_{1}\rangle_{F}$ and $|-\xi_{1}\rangle_{F}$ being the
coherent states of the oscillator with the amplitudes $\xi_{1}$ and
$-\xi_{1}$.
$|\psi_{A}^{+}\rangle=(|\uparrow\rangle_A+|\downarrow\rangle_A)/\sqrt{2}$
and
$|\psi_{A}^{-}\rangle=(|\uparrow\rangle_A-|\downarrow\rangle_A)/\sqrt{2}$
are the eigenstates of $\sigma_x$.

\begin{figure}
\center
  \includegraphics[width=0.9\columnwidth]{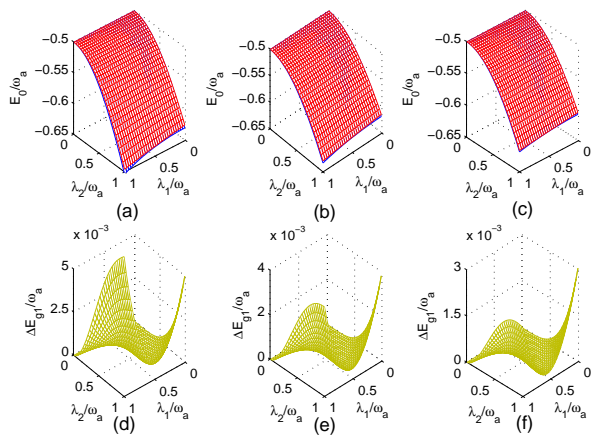} \caption{(Color
  online) The ground-state energy for the single-qubit ASRM obtained by
  the transformation method $E_0=E_{g1}$
  (red grid) and by the numerical
  solution $E_0=E_{g}$
  (blue grid) versus the coupling strengths
  $\lambda_{1}$ and $\lambda_{2}$: (a) $w_{b}=0.8w_{a}$; (b) $w_{b}=w_{a}$; (c)
  $w_{b}=1.2w_{a}$. The energy deviation $\Delta E_{g1}=E_{g1}-E_g$
  versus $\lambda_{1}$ and $\lambda_{2}$: (d) $w_{b}=0.8w_{a}$; (e) $w_{b}=w_{a}$; (f)
  $w_{b}=1.2w_{a}$.
  }\label{Fig.1.}
\end{figure}

The value of $\xi_{1}$ is obtained by numerically solving the
nonlinear equation (\ref{11}). $\xi_{1}$ has an approximately linear
dependence on the counterrotating-wave coupling strength by
neglecting high-order terms of the field mode as:
\begin{eqnarray}\label{29}
\xi_{1}&\simeq&\frac{\lambda_2}{2(w_a+w_b)}.
\end{eqnarray}
In Fig. 1, we compare the ground-state energy obtained by the
transformation method and that by the numerical solution.
Especially, we find that the ground-state energy obtained by the
transformation method coincides very well with the exactly numerical
solution when $|\lambda_1-\lambda_2|\leq0.15w_a$. Therefore, when
$\lambda_{1},\lambda_{2}\leq w_{a}$, the transformed ground-state
energy $E_{g1}$ approximates:
\begin{eqnarray}\label{15}
E_{g1}\simeq-\frac{1}{2}w_a-\frac{\lambda_2^2}{4(w_a+w_b)}+\frac{\lambda_2^3(\lambda_1-\lambda_2)}{8(w_a+w_b)^3},
\end{eqnarray}
which shows that the ground-state energy has an approximately
quadratic dependence on the coupling strength by neglecting
high-order terms of the field mode for the small factor
$|\lambda_1-\lambda_2|$ and is mainly contributed by the
counterrotating-wave interaction. This result differs further from
that of the SRM \cite{EPJD-59-473-2010}.

Considering the fidelity $F_{1}$ for the ground state
$|\phi_{g1}\rangle$, where $F_{1}=\langle\phi_{g1}|\phi_{g}\rangle$
and $|\phi_g\rangle$ is the ground state obtained through numerical
solutions \cite{arXiv-1303-3367v2-2013}, we plot $F_1$ as a function
of the coupling strengths $\lambda_{1}$ and $\lambda_{2}$ under
different detunings in Fig. 2. The result shows that the fidelity is
higher than $99.9\%$ when $\lambda_1\leq0.5w_{a}$ and
$\lambda_2\leq0.5w_{a}$. Furthermore, the fidelity under the
positive-detuning case ($w_{b}-w_{a}>0$) decreases slowest among all
the cases in Fig. 2 (a) - (c) when $\lambda_1$ and $\lambda_2$
increase.
\begin{figure}
\center
  \includegraphics[width=1\columnwidth]{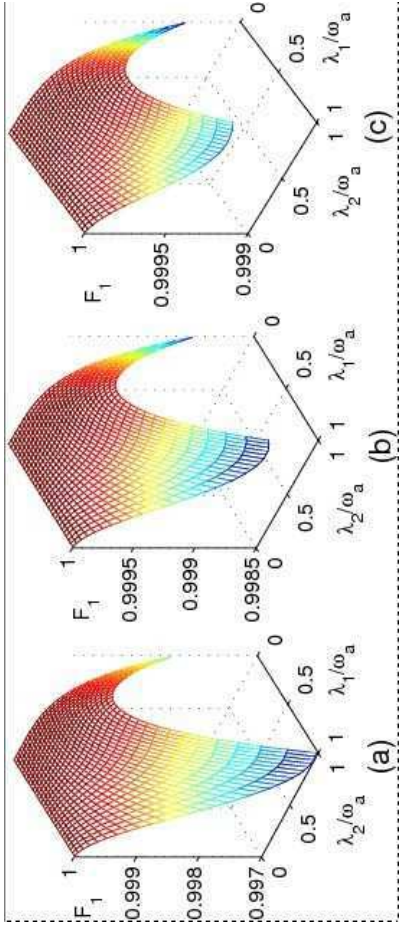} \caption{(Color
  online) The fidelity $F_{1}$ of the ground state for the single-qubit
  ASRM obtained by the transformation method
   versus the coupling strengths
  $\lambda_{1}$ and $\lambda_{2}$: (a) $w_{b}=0.8w_{a}$; (b) $w_{b}=w_{a}$; (c)
  $w_{b}=1.2w_{a}$.
  }\label{Fig.2.}
\end{figure}

\subsection{Ground-state entanglement}

In this section, we focus on the entanglement between the qubit and
the oscillator in the ground state of the single-qubit ASRM. Since
the ground state is a pure state, we take the von Neumann entropy as
an entanglement measure. If a pure state of a composite system $XY$
is given by the density matrix $\rho_{XY}$, the entropy of the
subsystem $X$ is defined as:
\begin{eqnarray}\label{SS}
S_{\rho_{X}} = -Tr(\rho_{X}log_2\rho_{X}),
\end{eqnarray}
where $\rho_{X}=Tr_{Y}(\rho_{XY})$ is the reduced density matrix for
the subsystem $X$ by tracing out the freedom degree of the subsystem
$Y$. Note that $S_{\rho_{X}}$ measures the entanglement between the
subsystems $X$ and $Y$ of the system, which has a maximum value of
$log_{2}K$ in a $K$-dimensional Hilbert space.

In the standard basis $\{|\uparrow\rangle_A,|\downarrow\rangle_A
\}$, the reduced density matrix of the qubit is
$\rho_{A}=Tr_{F}(|\Phi_{G}\rangle\langle\Phi_{G}|)$, where
$|\Phi_{G}\rangle$ is the exactly numerical ground state of the
single-qubit ASRM. The entropy of the qubit $S_{\rho_{A}}$ =
$-Tr(\rho_{A}log_2\rho_{A})$ is numerically plotted in Fig. 3, which
shows that the entanglement between the qubit and the oscillator
increases from as $\lambda_1$ and $\lambda_2$ increase from zero to
values close to $w_a$ and $w_b$.

\begin{figure}
\center
  \includegraphics[width=1\columnwidth]{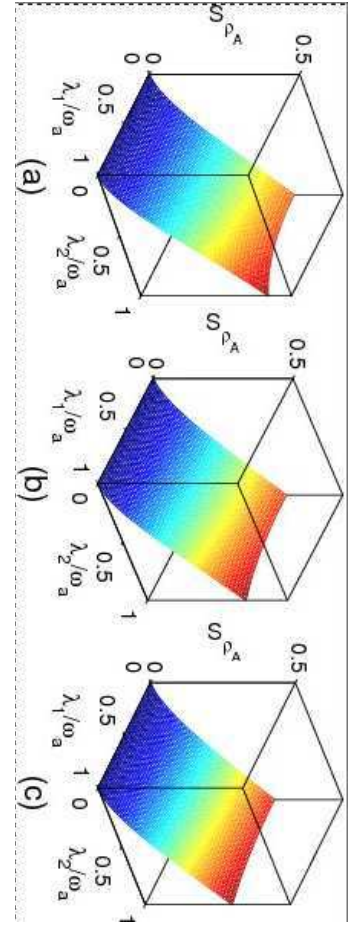} \caption{(Color
  online) The degree of entanglement $S_{\rho_A}$ for the qubit in the
  ground state of the single-qubit ASRM obtained by the numerical
  simulation versus the coupling strengths
  $\lambda_{1}$ and $\lambda_{2}$: (a) $w_{b}=0.8w_{a}$; (b) $w_{b}=w_{a}$; (c)
  $w_{b}=1.2w_{a}$.
  }\label{Fig.3.}
\end{figure}

\section{The two-qubit ASRM}

\subsection{Transformed ground state}

The Hamiltonian of the two-qubit ASRM is \cite{PRA-8-1440-1973}:
\begin{eqnarray}\label{17}
H_{}&=&w_{a}J_{z}+w_{b}b^{\dagger}b+g_{1}(b^{\dagger}J_{-}+bJ_{+})\cr\cr&&
+g_{2}(b^{\dagger}J_{+}+bJ_{-}),
\end{eqnarray}
where $w_{a}$ is the frequency of each qubit. $J_{l}\{
l=x,y,z,\pm\}$ describes the collective qubit operator of a spin-$1$
system. $b^{\dagger}$ ($b$) is the creation (annihilation) operator
of the harmonic oscillator with the frequency $w_{b}$. The
qubit-oscillator coupling strengths of the corotating-wave and
counterrotating-wave interactions are $g_{1}$ and $g_{2}$,
respectively. We denote the eigenstates of $J_{z}$ by
$|-1\rangle_{A}$, $|0\rangle_{A}$, and $|1\rangle_{A}$, i.e.,
$J_{z}|m\rangle_{A}=m|m\rangle_{A}$ ($m=0,\pm1$). $|0\rangle_{F}$ is
the vacuum state of the harmonic oscillator, and
$|\alpha\rangle_{F}$ denotes the coherent-state field with the
amplitude $\alpha$. When a rotation around the $y$ axis is
performed, the Hamiltonian of the two-qubit ASRM can be written as :
\begin{eqnarray}\label{18}
H_{2}&=&w_{a}J_{x}+w_{b}b^{\dagger}b+(g_{1}+g_{2})(b^{\dagger}+b)J_{z}\cr\cr&&
+i(g_{1}-g_{2})(b^{\dagger}-b)J_{y}.
\end{eqnarray}

To transform the Hamiltonian $H_{2}$ into a mathematical form
without counterrotating-wave terms, we apply a unitary
transformation to $H_{2}$:
\begin{eqnarray}\label{19}
H_{2}^{'}&=&e^{S_{2}}H_{2}e^{-S_{2}},
\end{eqnarray}
with
\begin{eqnarray}\label{20}
S_{2}&=&\xi_{2}(b^{\dagger}-b)J_{z},
\end{eqnarray}
where $\xi_{2}$ is a variable to be determined. Therefore, the
transformed Hamiltonian $H_{2}^{'}$ is decomposed into three parts:
\begin{eqnarray}\label{21}
H_{2}^{'}&=&H_{2}^{a}+H_{2}^{b}+H_{2}^{c},
\end{eqnarray}
with
\begin{eqnarray}\label{22-23-24}
H_{2}^{a}&=&w_{b}b^{\dagger}b+\bigg[
w_{a}\eta_{2}-(g_{1}-g_{2})\eta_{2}\xi_{2}\bigg]J_{x}\cr&&
+\bigg[w_{b}\xi_{2}^{2}-2\xi_{2}(g_{1}+g_{2}) \bigg]J_{z}^{2},\\
H_{2}^{b}&=&\bigg[
(g_{1}+g_{2})-w_{b}\xi_{2}\bigg](b^{\dagger}+b)J_{z}\cr&&+ i\bigg[
w_{b}\eta_{2}\xi_{2}+(g_{1}-g_{2})\eta_{2}\bigg](b^{\dagger}-b)J_{y},\\
H_{2}^{c}&=&w_{a}J_{x}\bigg\{
\cosh[\xi_{2}(b^{\dagger}-b)]-\eta_{2}\bigg\} \cr&&+
iw_{a}J_{y}\bigg\{
\sinh[\xi_{2}(b^{\dagger}-b)]-\eta_{2}\xi_{2}(b^{\dagger}-b)
\bigg\}\cr&&+ (g_{1}-g_{2})(b^{\dagger}-b)J_{x}\bigg\{
\sinh[\xi_{2}(b^{\dagger}-b)]\cr&&-\eta_{2}\xi_{2}(b^{\dagger}-b)
\bigg\}+i(g_{1}-g_{2})(b^{\dagger}-b)J_{y}\cr&&\times \bigg\{
\cosh[\xi_{2}(b^{\dagger}-b)]-\eta_{2}\bigg\}+
O(b^{\dagger2},b^{2}),
\end{eqnarray}
where $\eta_{2}=$ $_{F}\langle
0|\cosh[\xi_{2}(b^{\dagger}-b)]|0\rangle_{F}$ $=$
$e^{-\xi_{2}^{2}/2}$ and
$O(b^{\dagger2},b^{2})=(g_{1}-g_{2})\eta_{2}\xi_{2}J_{x}
(b^{\dagger2}-2b^{\dagger}b-b^{2})$. As shown in the single-qubit
ASRM, when $\xi_2$ and $|g_1\pm g_2|$ are much smaller than the
frequency sum $w_a+w_b$ where $w_a\approx w_b$, $H_{2}^{c}$ can be
neglected, thus $H_{2}^{'}\simeq H_{2}^{a}+H_{2}^{b}$. Compared with
$H_{1}^{a}$ in the single-qubit ASRM of Sec. II, the main difference
is the presence of the $J_{z}^{2}$ operator term in $H_{2}^{a}$, but
in the single-qubit ASRM the corresponding term $\sigma_z^2=1$ is
just a constant. Therefore, $H_{2}^{a}$ here represents a
renormalized three-level system in which we need to diagonalize
$H_{2}^{a}$ to remove counterrotating-wave terms.

The eigenvalues $\nu_{k}$ ($k=1,2,3$) and eigenstates
$|\varphi_{k}\rangle_{A}$ of the Hamiltonian
$H_{2}^{''}=H_{2}^{a}-w_{b}b^{\dagger}b$ are:
\begin{eqnarray}\label{25}
\nu_{1}&=&\frac{A}{2}-\frac{1}{2}\sqrt{A^{2}+8B^2}, \cr
|\varphi_{1}\rangle_{A}&=&\frac{1}{N_{1}}\bigg\{|-1\rangle_{A}-\frac{(A+\sqrt{A^2+8B^2})}{2B}|0\rangle_{A}+|1\rangle_{A}\bigg\},\cr\cr
\nu_{2}&=&A, \cr\cr
|\varphi_{2}\rangle_{A}&=&\frac{1}{N_{2}}\bigg\{-|-1\rangle_{A}+|1\rangle_{A}\bigg\},\cr\cr
\nu_{3}&=&\frac{A}{2}+\frac{1}{2}\sqrt{A^{2}+8B^2}, \cr\cr
|\varphi_{3}\rangle_{A}&=&\frac{1}{N_{3}}\bigg\{|-1\rangle_{A}-\frac{(A-\sqrt{A^2+8B^2})}{2B}|0\rangle_{A}+|1\rangle_{A}\bigg\},\cr&&
\end{eqnarray}
with
\begin{eqnarray}\label{26}
A&=&w_{b}\xi_{2}^2-2\xi_{2}(g_{1}+g_{2}), \cr
B&=&\frac{1}{\sqrt{2}}[w_{a}\eta_{2}-\eta_{2}\xi_{2}(g_{1}-g_{2})],
\end{eqnarray}
where $N_{k}$ is the normalization factor of the eigenvector
$|\varphi_{k}\rangle_{A}$. Here the eigenvalues are arranged in the
decreasing order: $\nu_{1}<\nu_{2}<\nu_{3}$. Then $H_{2}^{'}$ can be
expanded in terms of the renormalized eigenvectors:
\begin{eqnarray}\label{27}
H_{2}^{'}&\simeq&\sum_{k=1}^{3}\nu_{k}|\varphi_{k}\rangle_{A}\langle
\varphi_{k}|+\bigg[
(D_{1}b+D_{2}b^{\dagger})|\varphi_{1}\rangle_{A}\langle
\varphi_{2}|\cr&&+(D_{3}b+D_{4}b^{\dagger})|\varphi_{2}\rangle_{A}\langle
\varphi_{3}|+H.c.\bigg]+w_{b}b^{\dagger}b,
\end{eqnarray}
where $D_{x}\ (x=1,2,3,4)$ is the coefficient depending on the
variable $\xi_{2}$.

After transforming the Hamiltonian $H_{2}$ into $H_{2}^{'}$, we can
eliminate counterrotating-wave terms describing the coupling between
the lowest two eigenstates by setting:
\begin{eqnarray}\label{28}
D_{1}&=&\eta_{2}\bigg[
w_{a}\xi_{2}+(g_{1}-g_{2})\bigg]\bigg(A+\sqrt{A^2+8B^2}\bigg)\cr&&-2\sqrt{2}B\bigg[
(g_{1}+g_{2})-w_{b}\xi_{2}\bigg]=0.
\end{eqnarray}
The value of $\xi_{2}$ is obtained by numerically solving the
nonlinear equation (\ref{28}). We find that when $g_{1}\leq0.5w_{a}$
and $g_{2}\leq0.5w_{a}$, $\xi_{2}$ has an approximately linear
dependence on the coupling strengths:
\begin{eqnarray}\label{29}
\xi_{2}&\simeq&\frac{(w_{b}-w_{a})g_{1}+(w_{b}+w_{a})g_{2}}{w_{b}^2+w_{a}^2}.
\end{eqnarray}

\begin{figure}
\center
  \includegraphics[width=1\columnwidth]{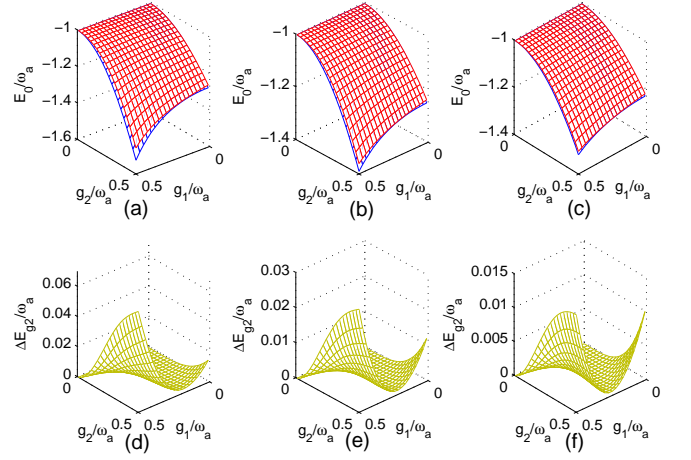} \caption{(Color
  online) The ground-state energy for the two-qubit ASRM obtained by the transformation
  method $E_0=E_{g2}$
  (red grid) and the numerical
  solution $E_0=E_{g}$
  (blue grid) versus the coupling strengths
  $g_{1}$ and $g_{2}$: (a) $w_{b}=0.8w_{a}$; (b) $w_{b}=w_{a}$; (c)
  $w_{b}=1.2w_{a}$, where $E_{g2}$ is plotted by using $\nu_{1}$ in Eq. (\ref{25}).
  The energy deviation $\Delta E_{g2}=E_{g2}-E_g$
  versus $g_{1}$ and $g_{2}$: (d) $w_{b}=0.8w_{a}$; (e) $w_{b}=w_{a}$; (f)
  $w_{b}=1.2w_{a}$.
  }\label{Fig.4.}
\end{figure}
In Fig. 4, we compare the ground-state energy obtained by the
transformation method and that obtained by the numerical solution.
We find that when $g_{1}\leq0.25w_{a}$ and $g_{2}\leq0.25w_{a}$, the
ground-state energy obtained by the transformation method coincides
very well with the exact value even for $|g_{1}-g_{2}|=0.24w_{a}$.
Therefore, when $g_{1}\leq0.5w_{a}$ and $g_{2}\leq0.5w_{a}$,
$|\varphi_{1}\rangle_{A}|0\rangle_{F}$ is expected to be the
approximately analytical ground state of the transformed
Hamiltonian, and the ground state $|\phi_{g}\rangle$ of the
two-qubit ASRM can be expressed by the transformed ground state
$|\phi_{g2}\rangle$:
\begin{eqnarray}\label{30}
|\phi_{g2}\rangle&=&e^{-S_{2}}|\varphi_{1}\rangle_{A}|0\rangle_{F}
\cr&&=\frac{1}{N_{1}}\big(
|-1\rangle_{A}|\xi_{2}\rangle_{F}-\frac{\nu_{3}}{B}|0\rangle_{A}|0\rangle_{F}+|1\rangle_{A}|-\xi_{2}\rangle_{F}\big),\cr&&
\end{eqnarray}
and the ground-state energy $E_{g2}$ is:
\begin{eqnarray}\label{31}
E_{g2}&\simeq&\nu_{1}\simeq
-w_{a}-\frac{(g_{1}+g_{2})g_{2}}{w_{a}w_{b}},
\end{eqnarray}
which directly shows that $E_{g2}$ has an approximately quadratic
dependence on the qubit-oscillator coupling strengths by neglecting
high-order terms of the field mode. This result differs further from
that in the two-qubit SRM \cite{arXiv-1303-3367v2-2013}.

\begin{figure}
\center
  \includegraphics[width=1\columnwidth]{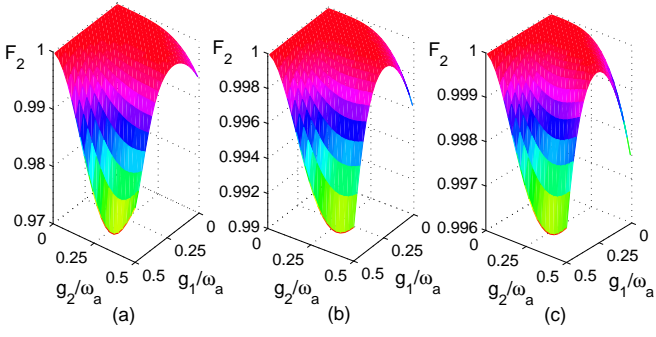} \caption{(Color
  online) The fidelity $F_{2}$ of the ground state for the two-qubit ASRM
  obtained by the transformation method
  versus the coupling strengths
  $g_{1}$ and $g_{2}$: (a) $w_{b}=0.8w_{a}$; (b) $w_{b}=w_{a}$; (c)
  $w_{b}=1.2w_{a}$.
  }\label{Fig.5.}
\end{figure}

The fidelity $F_{2}$ of the ground state as a function of the
qubit-oscillator coupling strengths $g_{1}$ and $g_{2}$ under
different detunings is plotted in Fig. 5. The result shows that
$F_{2}$ keeps higher than $99.9\%$ when $g_{1}\leq0.25w_{a}$ and
$g_{2}\leq0.25w_{a}$, which coincides with the behavior of the
transformed ground-state energy shown in Fig. 5.

\subsection{Ground-state entanglement}

We also examine the ground-state entanglement of the two-qubit ASRM
by taking into account both the transformation method and the
exactly numerical treatment. Negativity is taken to quantify the
entanglement for two qubits, which is defined as
\cite{PRA-65-032314-2002}:
\begin{eqnarray}\label{32}
M_{\rho_A}&=&\frac{\|\rho_A^{T}\|-1}{2},
\end{eqnarray}
where $\rho_A^{T}$ is the partially transposed matrix of the
two-qubit reduced density matrix $\rho_A$, with
$\rho_A=Tr_{F}(\rho_{AF})$ and $\rho_{AF}=|\phi_{g}\rangle\langle
\phi_{g}|$, and $\|\rho_A^{T}\|$ is the trace norm of $\rho_A^{T}$.
Thus, $M_{\rho_A}$ alternatively equals the absolute value for the
sum of the negative eigenvalues of $\rho_A^{T}$. For the transformed
ground state $|\phi_{g2}\rangle$ in Eq. (\ref{30}), the partially
transposed matrix of the reduced density operator for the two qubits
in the qubit basis $\Gamma_{q}$ $=$ $\{$
$|\uparrow_{1}\rangle|\uparrow_{2}\rangle,
|\uparrow_{1}\rangle|\downarrow_{2}\rangle,
|\downarrow_{1}\rangle|\uparrow_{2}\rangle,
|\downarrow_{1}\rangle|\downarrow_{2}\rangle$ $\}$, where
$|\uparrow_{l}\rangle$ and $|\downarrow_{l}\rangle$ ($l=1,2$)
correspond to the excited and ground states of the $l$th qubit
respectively, is obtained as follows:
\begin{figure}
\center
  \includegraphics[width=0.95\columnwidth]{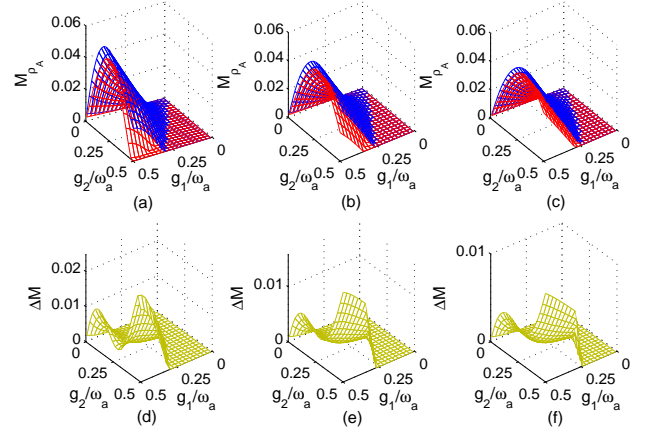} \caption{(Color
  online) The negativity $M_{\rho_A}$ of two qubits in the ground state of the
  two-qubit ASRM
  versus the coupling strengths
  $g_{1}$ and $g_{2}$: (a) $w_{b}=0.8w_{a}$; (b) $w_{b}=w_{a}$; (c)
  $w_{b}=1.2w_{a}$. The results obtained by the
  transformation method $M_{\rho_A}=M_{\rho_A}^{t}$ and the numerical simulation
  $M_{\rho_A}=M_{\rho_A}^{n}$ are
  represented by the red grid and the blue grid, respectively. The deviation in
  the two-qubit
  negativity $\Delta M=M_{\rho_A}^{n}-M_{\rho_A}^{t}$ obtained by
  transformation method: (d) $w_{b}=0.8w_{a}$;
  (e) $w_{b}=w_{a}$; (f) $w_{b}=1.2w_{a}$.
  }\label{Fig.6.}
\end{figure}
\begin{eqnarray}\label{33}
\rho_A^{T}&=&\frac{1}{(2+\beta^2)} \left(\begin{array}{cccc}
1 & \frac{\beta}{\sqrt{2}}e^{-\frac{\alpha^2}{2}} & \frac{\beta}{\sqrt{2}}e^{-\frac{\alpha^2}{2}} & \frac{\beta^2}{2} \\
\frac{\beta}{\sqrt{2}}e^{-\frac{\alpha^2}{2}} & \frac{\beta^2}{2} & e^{-2\alpha^2} & \frac{\beta}{\sqrt{2}}e^{-\frac{\alpha^2}{2}}\\
\frac{\beta}{\sqrt{2}}e^{-\frac{\alpha^2}{2}} & e^{-2\alpha^2} & \frac{\beta^2}{2} & \frac{\beta}{\sqrt{2}}e^{-\frac{\alpha^2}{2}} \\
\frac{\beta^2}{2} & \frac{\beta}{\sqrt{2}}e^{-\frac{\alpha^2}{2}} &
\frac{\beta}{\sqrt{2}}e^{-\frac{\alpha^2}{2}} & 1
\end{array}
\right),\cr&&
\end{eqnarray}
where $\alpha=\xi_{2}$ and $\beta=-\frac{\nu_{3}}{B}$. With Eq.
(\ref{33}), we can calculate the negative $M_{\rho_A}$:
\begin{eqnarray}\label{34}
M_{\rho_A}&=&\max \bigg\{
\frac{2e^{-2\xi_{2}^2}-(\frac{\nu_{3}}{B})^2}{2[2+(\frac{\nu_{3}}{B})^2]},
0 \bigg\}.
\end{eqnarray}
When $g_1\leq0.25w_{a}$ and $g_2\leq0.25w_{a}$, $M_{\rho_A}$
approximates:
\begin{eqnarray}\label{35}
M_{\rho_A}&\simeq&
\frac{w_{b}\big[(1-\frac{1}{\sqrt{2}})^2g_{2}^{2}+g_1g_2\big]}{4w_{a}(w_{a}+w_{b})^2}.
\end{eqnarray}
From Eq. (\ref{35}), we see that the two-qubit entanglement
increases with $g_2^2$ and $g_1g_2$. The two-qubit negativity as a
function of the qubit-oscillator coupling strengths $g_{1}$ and
$g_{2}$ under different detunings has been plotted in Fig. 6 (a) -
(c), and the corresponding deviation from the numerical simulation
is plotted in Fig. 6 (d) - (f). For $0<g_1\leq0.25w_a$ and
$0<g_2\leq0.25w_a$, the two-qubit negativity has a linear dependence
on $g_1$ for the fixed $g_2$ and a quadratic dependence on $g_2$ for
the fixed $g_1$; For $0<g_1\leq0.25w_a$ and $0.25w_a<g_2<0.5w_a$,
the negativity keeps close to zero; However, for
$0.25w_a<g_1<0.5w_a$ and $0<g_2<0.5w_a$, the negative has a similar
dependence on $g_1$ and $g_2$ with the case of $0<g_1\leq0.25w_a$
and $0<g_2\leq0.25w_a$. We find that when $g_1\leq0.25w_{a}$ and
$g_2\leq0.25w_{a}$ the deviation in the negativity is close to zero,
meaning the ground state obtained by the transformation method
agrees well with the exact one. This directly shows that the
two-qubit entanglement is caused by the counterrotating-wave
interaction in the Hamiltonian. Interestingly, after the negativity
has reached its maximum, it will monotonically decrease when $g_{1}$
or $g_{2}$ further increases. Furthermore, the maximum of the
two-qubit entanglement in the two-qubit ASRM is far larger than that
in the two-qubit SRM, and the two-qubit entanglement mainly appears
when the coupling strength of the corotating-wave interaction is
bigger than that of the counterrotating-wave interaction, which is
because the contribution to the two-qubit entanglement from the
counterrotating-wave interaction is larger than that from the
corotating-wave interaction in Eq. (\ref{35}). As seen from Fig. 7,
when $g_{1}>1.11w_{a}$ or $g_{2}>0.88w_{a}$ at $w_{b}=w_{a}$,
$M_{\rho_A}$ decreases to zero and never increases again, and the
maximum negativity is about $0.10$ which is only $3.5\times10^{-2}$
in the two-qubit SRM \cite{arXiv-1303-3367v2-2013}.
\begin{figure}
  \includegraphics[width=0.5\columnwidth]{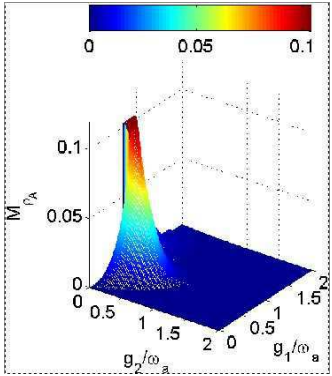} \caption{(Color
  online) The negativity $M_{\rho_A}$ of two qubits in the ground state of the two-qubit
  ASRM obtained by the numerical simulation
  versus the coupling strengths
  $g_{1}$ and $g_{2}$ when $w_{b}=w_{a}$.
  }\label{Fig.7.}
\end{figure}

In Fig. 8, we numerically plot the entropy $S_{\rho_A}$ of two
qubits versus the coupling strengths $g_1$ and $g_2$ in the ground
state of the two-qubit ASRM, where $S_{\rho_A}$ $=$
$-Tr(\rho_{A}log_2\rho_{A})$. The result shows that the entanglement
between the qubit and the oscillator increases from as $g_1$ and
$g_2$ increase from zero to values close to $w_a$ and $w_b$.

\begin{figure}
\center
  \includegraphics[width=1\columnwidth]{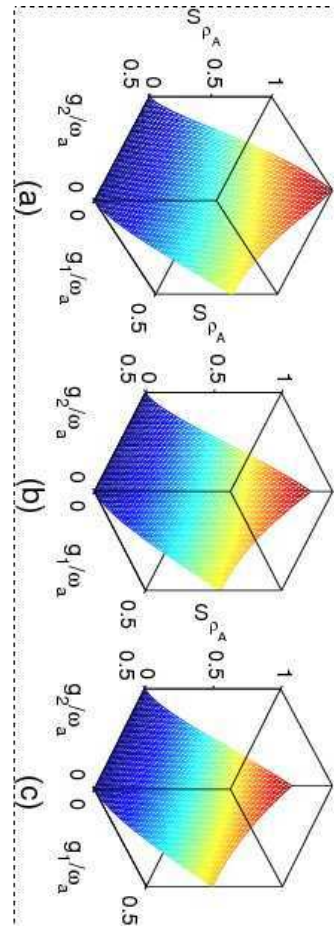} \caption{(Color
  online) The degree of entanglement $S_{\rho_A}$ for the qubits in the
  ground state of the two-qubit ASRM obtained by numerical
  simulations versus the coupling strengths
  $g_{1}$ and $g_{2}$: (a) $w_{b}=0.8w_{a}$; (b) $w_{b}=w_{a}$; (c)
  $w_{b}=1.2w_{a}$.
  }\label{Fig.8.}
\end{figure}

\section{Conclusion}

In conclusion, we have used the transformation method to obtain the
approximately analytical ground states of the single- and two-qubit
ASRMs, and shown that the transformed results coincided well with
those obtained by numerical simulations for a wide range of
parameters under the near-resonance condition. We find that the
ground-state energy in either the single- or two-qubit ASRM has an
approximately quadratic dependence on the qubit-oscillator coupling
strengths, and the contribution of the counterrotating-wave
interaction on the ground state energy is larger than that of the
corotating-wave interaction. Interestingly, we also find that the
two-qubit entanglement of the two-qubit ASRM decreases to zero and
never increases again as long as the qubit-oscillator coupling
strengths are large enough. Furthermore, the maximum of the
two-qubit entanglement in the two-qubit ASRM is far larger than that
in the two-qubit SRM, and the two-qubit entanglement mainly appears
when the coupling strength of the corotating-wave interaction is
bigger than that of the counterrotating-wave interaction.

\section{Acknowledgement}

This work is supported by the Major State Basic Research Development
Program of China under Grant No. 2012CB921601, the National Natural
Science Foundation of China under Grant No. 11374054, No. 11305037,
No. 11347114, and No. 11247283, the Natural Science Foundation of
Fujian Province under Grant No. 2013J01012, and  the funds from
Fuzhou University under Grant No. 022513, Grant No. 022408, and
Grant No. 600891.

%%%%three

\end{document}